# interface_master: Python package building CSL and approximate CSL interfaces of any two lattices – an effective tool for interface engineers.

**Yaoshu Xie[a]\*, Kiyou Shibata[a] and Teruyasu Miozguchi[a]\***

[a]Institute of Industrial Science, the University of Tokyo, 4-6-1, Meguro-ku, Tokyo, 13-8505, Japan

Correspondence email: ysxie@iis.u-tokyo.ac.jp; teru@iis.u-tokyo.ac.jp

**Abstract** Crystalline interfaces are of highly importance in many practical applications. To conduct effective simulation and analysis for coincident site lattice (CSL) interfaces, effective programmes are of high demand in building their CSL bicrystal models to apply periodic boundary condition. The existing reported programmes capable suffer from problems including limitation of available lattice, poor capability for making heterogeneous interfaces and requirement of none-intuitive input parameters which cannot be directly obtained from an experimental observation result for an interface. This work presents a python programme to construct bicrystal model of interfaces consisting of any two lattices, including tilt, twist, mixed, and hetero interfaces and surfaces, which will support in effective simulation and analysis of crystalline interfaces.

**Program Summary**
**Program title: interface_master**
This program aims to construct bicrystal model of interfaces consisting of any two lattices which will support in effective simulation and analysis of interfaces.
**Licensing provisions:** MIT license (MIT)
**Programming language:** Python

**Nature of problem:** Atomistic computational simulation is an effective way to study crystalline interfaces where one often needs to construct a coincident site lattice (CSL) interface to apply periodic boundary condition. While there have been reported packages for building CSL interfaces[1–4], none of them are capable to build an arbitrary CSL interface by simply knowing the geometric information of the disorientation of the two crystals forming this interface and the interface plane orientation. The main difficulty in making a general program to do so is to compute the approximate CSL which is to slightly deform one of the two crystals to make them three-dimensionally or two-dimensionally coincident given an arbitrary disorientation.

**Solution method:** In this work, applying two effective algorithms[5–7] we generated a python programme capable to build both CSL and approximate CSL interfaces. By inputting simply two CIF files (better of conventional cells), and some intuitive geometric parameters, users can build any desired CSL interface.

1. **Introduction**

Crystalline interfaces play significant roles in modern materials science. Homogeneous crystalline interfaces known as grain boundaries (GBs) can dominate the performances of many important engineering materials existing as polycrystallines which has bred the concept of GB engineering to promote the performance of polycrystalline materials by tailoring the distribution of distinguishable GBs [8,9]. On the other hand, heterogeneous crystalline interfaces are ubiquitous in modern functional devices and play key roles in device performances. Examples of these significant interfaces include not only those in the devices which are already commercialized like electrode/electrolyte interface in solid-state lithium batteries[10], and the semiconductor/electrode interface in thin-film transistors[11]. Exploring the interface physics like interface superconductivity[12] is also a key aspect to achieve well-controlled novel quantum materials.

Computational simulation is an effective way to study the interfaces. In most cases, one needs to construct a coincident site lattice (CSL) interface for simulation. A CSL interface comprises two coincident crystalline surfaces and it is desired by most atomic simulation methods to allow use of two-dimensional periodic boundary condition in the interface plane. Of course, CSL interface is not simply a mathematical concept supporting in simulation and it is of practical importance. CSL GBs exist in many important engineering materials and the main objective of GB engineering is often to increase the composition of certain CSL GB to achieve desired performance. CSL heterogeneous interfaces are the most important structures for all the epitaxial thin-film devices where the substrate and deposited film were desired to keep coincident for a good thin-film performance.

Exploration of CSL interfaces can be done in two ways. Firstly, one can apply numerical searching for any two lattices to find two coincident surfaces. Secondly, one can find a CSL interface from the geometric information by the overlapping diffraction patterns of the two crystals forming the interface. Among all the reported computer programs for constructing CSL interfaces, only one named MPInterfaces[3] applying a lattice-matching method is capable to explore CSL interfaces of any two materials and is available for those combined by crystals of any the 14 Bravais lattices. However, although it provides an effective tool to determine the interface structure by high-throughput simulation, it does not provide a straightforward usage to construct a specific CSL interface given its geometric information including the disorientation of the two lattices and the interface plane orientation (five-dimensional macroscopic parameters). This usage is demanding both for engineers who desire to simulate the interfaces observed by experiments without conducting crystallographic computation manually, and for one needing to explore the structure-property relationship of the interfaces in the five-dimensional space of macroscopic parameters. Some previous studies have been conducted to investigate the structure-property relationship in this five-dimensional space for GBs of cubic lattice systems[13,14] while such studies are still rare for interfaces involving non-cubic crystals due to lacking in effective code to construct interface models. Considering the general cases to simulate interfaces, given two arbitrary crystals defined by two cif files and an arbitrary disorientation which can be measured from experiment or pre-determined by sampling a point in the 5-dimensional space, one often needs to slightly deform one of the two crystals to make them coincident forming a three-dimensional or two-dimensional CSL. However, none of the previously reported programmes provide this application. Another shortage of the MPInterfaces package is that it cannot be applied to search for three-dimensional CSL. For those applying periodic boundary condition in all the three dimensions of the supercell, there always exist two interfaces in the supercell. In this case, applying a supercell consisting of two three-dimensionally coincident crystals where the two crystals coincide not only in the interface plane but also in the direction crossing the interface is a convenient way to make the two interfaces identical. Such convention has been applied in many first-principles simulation for symmetric tilt

GBs[15,16] of cubic lattices. Lacking in capability to automatically build such three-dimensional CSL interfaces for non-cubic lattice systems has become a main limitation to extend such studies to more general cases. Here, we provide a python package which is capable to achieve a convenient usage satisfying all these purposes above. This package is capable to make approximate CSL interfaces of any two crystalline materials by inputting two Cif files and geometric parameters with good intuition. This code outputs a structure file of the constructed interface with format as VASP or LAMMPS data file, as well as generates a 'Structure' object of this interface corresponding to usages for packages like Pymatgen[17] and Atomic Simulation Environment[18]. It also provides computation of cell of non-identical displacement (CNID) of the constructed CSL interface to accelerate the exploration of interface structures by rigid body translation (RBT). Although we have published a brute-force method for computing CNID [19], in this code we applied an analytical algorithm which is more effective and reliable. In this article we will briefly review the most important algorithms, and present the application of this code in building some experimentally observed interfaces. This package is now available from github: https://github.com/nmdl-mizo/interface_master/tree/master.

## 2. Algorithms
### 2.1 Algorithm generating approximate CSL for any two lattices

Here we present the algorithm finding an approximate CSL for any two lattices with certain initial disorientation. The algorithm was proposed by Bonnet et al.[5–7] and we did some modifications. Given two lattices 1 and 2 expressed by basis vectors $a_i^1$ and $a_i^{2,0}$ ($i$ =1,2,3) as column vectors, where $a_i^{2,0}$ is related to basis vectors of rotated lattice 2 $a_i^2$ by a rotation operation $R$, denominate an operation $U$ relating $a_i^1$ and $a_i^2$ as:

$$(a_i^1)U = (a_i^2) = R(a_i^{2,0}), \qquad 2.1.1$$

where $(a_i^j) = [a_1^j \ a_2^j \ a_3^j]$ is a 3×3 matrix representing the set of basis vectors of lattice $j$.

If lattice 1 and 2 form an approximate CSL with basis vector sets $(x_i)$, we have

$$(x_i) = (a_i^1)U_1 = R(D(a_i^{2,0}))U_2, \qquad 2.1.2$$

where $D$ slightly deforms $a_i^{2,0}$ and can be decomposed as $D = I + \delta D$ using $\delta D$ with small enough eigenvalues; $U_1$ and $U_2$ execute linear combinations extracting supercells of $(a_i^1)$ and $RD(a_i^{2,0})$. $U_1$ and $U_2$ have only integer elements. Denominate the ratio of the volume of $(x_i)$ to $(a_i^1)$, $(Da_i^2)$ as $\Sigma_1 = \det(U_1)$, $\Sigma_2 = \det(U_2)$.

Note that in the original literature by Bonnet et al.[5–7], $U_1 \ U_2 \ U$ were all applied as left-operations, while we found that their following discussions indicated that they should all be right-operations if the basic vectors were column vectors. *(Applying a right-operation supports the fact that $U_1$, $U_2$ have only integer elements because CSL vectors should be linear combination of lattice 1 & 2 both with only* integer coefficients. On the other hand, the equivalent left-operation is a linear transformation with $A = (a_i^1)U(a_i^1)^{-1}$ and can have non-integer elements.)

To find the relationship between the approximate CSL and the two original lattices, rewrite 2.1.2 as

$$(a_i^1)U' = R'(a_i^{2,0})$$
$$U' = U_1 U_2^{-1}$$
$$R' = RD$$

2.1.3

Comparing 2.1.3 with 2.1.1, $U'$, $R'$ should be close to $U$, $R$. Therefore, the process finding an approximate CSL of lattice 1 and 2 is to find a rational matrix $U'$ close to $U$ so that the matrix $D$ to hold 2.1.3 is close to a unit matrix. Note that $U'$ is a rational matrix and its elements can be written as $u'_{ij}/N$ where at least one of the nine $u'_{ij}$s is prime to $N$. To control the extend of approximation, defining parameters $\Delta u$, $\Delta d$ and denominating elements of $D, U, I$ (a unit matrix) as $d_{ij}$, $u_{ij}$, $\delta_{ij}$ we make some inequalities:

$$\max_i \sum_j |d_{ij} - \delta_{ij}| \leq \Delta d$$
$$\max_{i,j} |u'_{ij}/N - u_{ij}| \leq \Delta u$$
$$\det(U_1) \leq \Sigma_{1max}; \det(U_2) \leq \Sigma_{2max}$$

2.1.4

As $u'_{ij}/N \approx u_{ij}$, $u'_{ij}$ can be determined by

$$u'_{ij} = round(Nu_{ij})$$

2.1.5

where $N \leq \Sigma_{2max}$.

Now we present the algorithm.
**Input** $(a_i^1)$, $(a_i^{2,0})$, rotation axis $r$, angle $\theta_0$, $\Sigma_{1max}$, $\Sigma_{2max}$, $\Delta u$, $\Delta \theta$, and $d\theta$;
$\theta = \theta_0$
while $\theta < \theta_0 + \Delta\theta$:
    $R \leftarrow r, \theta$
    $N = 1$
    while $N \leq \Sigma_{2max}$:
        compute $U'$ by 2.1.5
        if $|u'_{ij}/N - u_{ij}| \leq \Delta u$:
            compute $D$ by 2.1.3
            if $\max_i \sum_j |d_{ij} - \delta_{ij}| \leq \Delta d$
                compute $U_1$ and $U_2$ using algorithm in **2.2**
                if $\det(U_1) \leq \Sigma_{1max}$:
                      **Output** $U_1$, $U_2$ $D$ and $R$.

```
            end if
        end if
    end if
    N += 1
  end while
  θ += dθ
end while
```

In more general cases the two crystals can form a 2-D CSL without simultaneously forming a 3-D CSL. Our programme is also available for this case. The process is as follows.

***Input:*** $(a_i^1)$, $(a_i^{2,0})$, (h$_1$ j$_1$ k$_1$), (h$_2$ j$_2$ k$_2$), a target disorientation $\boldsymbol{R^t}$.

1) Rotate $(a_i^{2,0})$ by $\boldsymbol{R^*}$ so that the normal vector of (h$_1$ j$_1$ k$_1$) $\boldsymbol{n_1}$ is collinear with the normal vector of (h$_2$ j$_2$ k$_2$) $\boldsymbol{n_2}$;

2) Rotate $(a_i^{2,0})$ along the (h$_1$ j$_1$ k$_1$) by $\boldsymbol{R'}$ so that $\boldsymbol{R'R^*} \approx \boldsymbol{R^t}$;

3) Applying the method by Banadaki & Patala[20] find two primitive plane bases of (h$_1$ j$_1$ k$_1$) and (h$_2$ j$_2$ k$_2$), denominated as [**a₁ b₁**], [**a₂ b₂**].

4) Define two lattices $(a_i^{1*})$, $(a_i^{2,0*})$ as [**a₁ b₁ a₁×b₁**], [**a₂ b₂ a₁×b₁**].

5) Execute the algorithm searching for the 3-D approximate CSL lattice with substituting $(a_i^1)$, $(a_i^{2,0})$, **r** by $(a_i^{1*})$, $(a_i^{2,0*})$, $\boldsymbol{n_1}$ with keeping the third column of the matrix $u'_{ij}/N$ always being [0, 0, 1].

***Output:*** $\boldsymbol{U_1}$, $\boldsymbol{U_2}$, $\boldsymbol{D}$, $\boldsymbol{R}$. The two vectors of $(a_i^{1*})\boldsymbol{U_1}$ or $\boldsymbol{RD}(a_i^{2,0*})$ perpendicular to $\boldsymbol{n_1}$ form a primitive basis of the 2-D CSL lattice.

## 2.2 Algorithm generating a 3-D DSC basis and a 3-D CSL basis for two arbitrary coincident lattices.

Here we review the method to compute $\boldsymbol{U_1}$, $\boldsymbol{U_2}$ mentioned in **2.1** and to compute CNID (two-dimensional DSC) and CSL[6]. Considering two arbitrary lattices with basic vectors as $(a_i^1)$ and $(a_i^2)$ forming an exact CSL lattice $x_i$, we have

$$x_i = (a_i^1)U_1 = (a_i^2)U_2$$
$$(a_i^2) = (a_i^1)U$$
$$U = U_1(U_2)^{-1}$$

Similar with that in **2.1**, $U_1$ and $U_2$ have only integer elements with $\det(U_1) = \Sigma_1$, $\det(U_2) = \Sigma_2$, and elements of $U$ have can be written as $u_{ij}/N$, where $N \leq \Sigma_2$.

As lattice 1 is a sub-lattice of DSC [21], a basis of DSC $\boldsymbol{d_i}$ can be parameterized as:

$$\boldsymbol{d_1} = \frac{1}{\lambda}\boldsymbol{a_1^1},$$

$$d_2 = \frac{\alpha}{\mu}d_1 + \frac{1}{\mu}a_2^1,$$

$$d_3 = \frac{\beta}{\nu}d_1^1 + \frac{\gamma}{\nu}d_2^1 + \frac{1}{\nu}a_3^1,$$

where $\alpha, \beta, \gamma, \lambda, \mu, \nu$ are all non-negative integers.

To verify that such a DSC basis is primitive, the six non-negative integer parameters $\alpha, \beta, \gamma, \lambda, \mu, \nu$ should satisfy that:

$$\begin{cases} 0 \leq \alpha < \mu \\ 0 \leq \beta < \nu \\ 0 \leq \gamma < \nu \end{cases}$$

Then, the DSC bases can be determined by find the six integer parameters. The method solving the integer equations of the six parameters is presented in [6]. In this way, we obtain a DSC basis here.

As discussed in **2.1**, we need to compute $U_1$ and $U_2$ to find a CSL basis. According to the literature, a CSL basis can be computed by repeating the routine computing a DSC basis of an 'auxiliary' lattice $(a_i^{2a})$. The process is

1) Denominate a matrix $U_a$ related to $U$ by a symmetry operation along the second diagonal:

$$U_a = \frac{1}{N}\begin{pmatrix} u_{33} & u_{23} & u_{13} \\ u_{32} & u_{22} & u_{12} \\ u_{31} & u_{21} & u_{11} \end{pmatrix}$$

2) For $(a_i^1)$ and the 'auxiliary' lattice $(a_i^{2a})$ where $(a_i^{2a}) = (a_i^1)U_a$, using the method above compute a DSC basis $H_1$.

3) $H_1$ expressing $(a_i^{2a})$ as $(a_i^{2a}) = H_1 P$, where $P$ has elements as $p_{ij}$, and $U_1$ is determined as:

$$U_1 = \begin{pmatrix} p_{33} & p_{23} & p_{13} \\ p_{32} & p_{22} & p_{12} \\ p_{31} & p_{21} & p_{11} \end{pmatrix}$$

Then $U_2 = U^{-1}U_1$

## 2.3 Computation of CNID

After obtaining a CSL basis, we can specify an interface lying in a CSL lattice plane. Then the CNID of this interface can be found the following routine:

1) Specify an interface with miller indices (hkl)$_c$ expressed in the CSL lattice's frame;
2) Use the method in [19] to convert (hkl)$_c$ to be expressed in lattice 1's & lattice 2's frame as (hkl)$_1$ & (hkl)$_2$;
3) Use the method in [20] to find the two plane bases of (hkl)$_1$ & (hkl)$_2$ denominated as $[t_1^1 \; t_2^1]$ & $[t_1^2 \; t_2^2]$;
4) Compute a DSC basis of $[t_1^1 \; t_2^1 \; t_1^1 \times t_2^1]$ & $[t_1^2 \; t_2^2 \; t_1^1 \times t_2^1]$. Then the two of the basic DSC vectors lying in the interface is the desired CNID.

Bonnet's algorithm is both effective and generally available. It is effective because the only numerical searching aspect involved is to solve two equations under inspection in a small range of unknown integer parameters. It is generally available because once two lattices form an exact CSL lattice, this algorithm can be applied for them, no matter what lattice species the two lattices are and no matter whether they are identical or not.

## 3. Generation of bicrystal model and examples of application

The Cif files of the crystals making these example interfaces were provided by Materials Project [22]. Visualization is down using VESTA[23] and OVITO[24].

### 3.1 Generating non-cubic CSL GBs with known geometric information

Here we show an example of $CuInSe_2$ twinning GB [25] generated using our code. In this example, we input a cif file of a conventional cell of $CuInSe_2$, and output a bicrystal of a well-known twinning structure of this material as showed in the Figure 1.

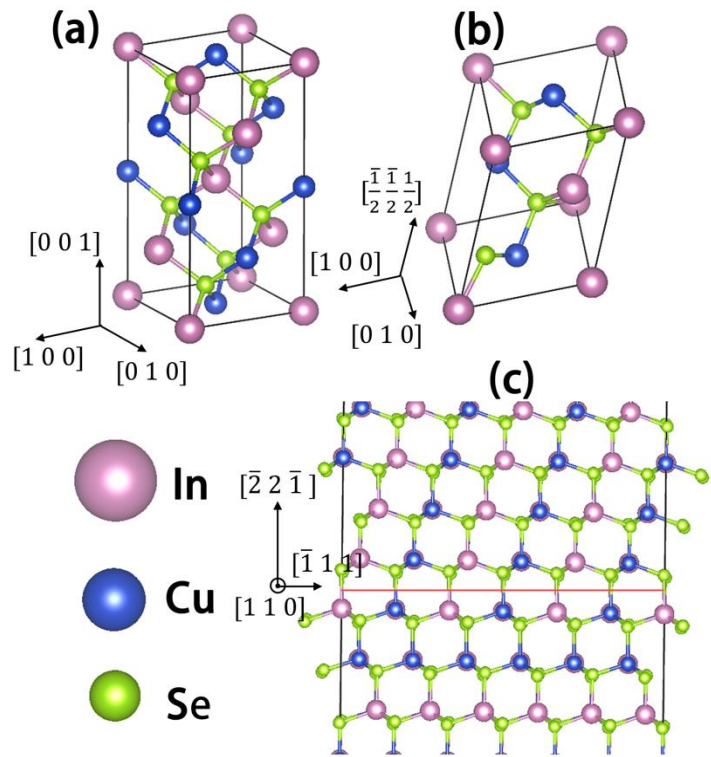

Figure 1 Unit cell and twinning GB of **CuInSe$_2$**; all the vectors expressed in the conventional cell's frame. (a) conventional cell; (b) primitive cell; (c) part of the final output bicrystal.

The process generating this GB is: 1) read the cif file of the conventional cell; 2) obtain the primitive cell and generate two identical lattices as $a_i^1$ and $a_i^{2,0}$; 3) according to the known geometry information, rotate $a_i^{2,0}$ to a close orientation (in this case rotate along axis [2 $\bar{2}$ 1] by 180 degrees); 4) near the orientation, using the algorithm in **2.1** find an approximate 3-D CSL and determine parameters $U, U_1, U_2, R, D$; 5) specify the interface to lie in a CSL lattice plane (hkl)$_c$; 6) find a CSL plane basis [y$_2$ y$_3$] of (hkl)$_c$ using the method in [20]; 7) find a CSL primitive basic vector **y$_1$** which is closely normal to the (hkl)$_c$ plane; 8) get the integer indices matrix $U_b^1$ so that $a_i^1 U_b^1 = [y_1\ y_2\ y_3]$ and $U_b^2 = U^{-1} U_b^1$; 9) generate two slabs with basis as column vectors of $a_i^1 U_b^1$ and $RD a_i^{2,0} U_b^2$; 10) translate the second slab by x$y_1$ and the GB is generated. Some of the mentioned parameters of practical interests are shown in Table 1.

As can be seen, because of some slight difference between the structure of a real material observed in the experiment and the structure provided by a cif file, the original crystals often cannot form an exact CSL by inputting geometry information obtained from experiment, which means that this CSL interfaces are not capable to be built with the existing programs without manual computation. For example, in Pymatgen, users need to adjust the lattice shape by their own to make this CSL structure while here we help users to find this deformation information automatically within allowed approximation. Our program also automatically computes the CNID cell which is output as two vectors (in this example as $a_i^1 [0\ \bar{1}\ \bar{1}]^T$ and $a_i^1 [\bar{1}\ \bar{1}\ 0]^T$).

Table 1 Parameters involved generating the CuInSe$_2$ twinning GB

| $a_i^1, a_i^{2,0}$ | $U_1$ | $U_2$ | $D$ | | |
|---|---|---|---|---|---|
| $2.8905 \begin{bmatrix} 2 & 0 & \bar{1} \\ 0 & 2 & \bar{1} \\ 0 & 0 & 2 \end{bmatrix}$ | $\begin{bmatrix} 0 & 1 & \bar{1} \\ 1 & 1 & 1 \\ 1 & 0 & \bar{1} \end{bmatrix}$ | $\begin{bmatrix} 0 & \bar{1} & \bar{2} \\ \bar{1} & \bar{1} & 0 \\ \bar{1} & 0 & \bar{1} \end{bmatrix}$ | $\begin{bmatrix} 1.0031\ldots & -0.0031\ldots & -0.0061\ldots \\ -0.0031\ldots & 1.0031\ldots & 0.0061\ldots \\ 0.0031\ldots & -0.0031\ldots & 0.99381\ldots \end{bmatrix}$ | | |
| $U_b^1$ | $U_b^2$ | $\theta, r$ | (hkl), (hkl)$_c$ | | |
| $\begin{bmatrix} \bar{3} & 0 & 1 \\ 1 & 1 & 1 \\ \bar{2} & 1 & 0 \end{bmatrix}$ | $\begin{bmatrix} 3 & 0 & \bar{1} \\ 1 & \bar{1} & \bar{1} \\ \bar{2} & \bar{1} & 0 \end{bmatrix}$ | 180°, [2 $\bar{2}$ 1] | (1 $\bar{1}$ 1), (0 0 1) | | |

Table 2 Parameters involved generating the Tellurium twinning GB

| $a_i^1, a_i^{2,0}$ | $U_1$ | $D$ | $U_2$ |
|---|---|---|---|
| $\begin{bmatrix} -a & -a & 0 \\ -b & b & 0 \\ 0 & 0 & c \end{bmatrix}$<br>$a = 2.25618\ \text{Å}$<br>$b = 3.90783\ \text{Å}$<br>$c = 5.95989\ \text{Å}$ | $\begin{bmatrix} 1 & 1 & 3 \\ 2 & 0 & \bar{1} \\ 0 & 1 & \bar{2} \end{bmatrix}$ | $\begin{bmatrix} 0.998\ldots & 0 & 0.0011\ldots \\ 0 & 1 & 0 \\ 0.0019\ldots & 0 & 1.0014\ldots \end{bmatrix}$ | $\begin{bmatrix} \bar{1} & \bar{1} & 4 \\ 1 & \bar{1} & 3 \\ 0 & 1 & 2 \end{bmatrix}$ |
| $U_b^1$ | $U_b^2$ | $R^t$ | (hkl), (hkl)$_c$ |
| $\begin{bmatrix} 4 & 1 & 1 \\ 1 & 2 & 0 \\ \bar{2} & 0 & 1 \end{bmatrix}$ | $\begin{bmatrix} 3 & \bar{1} & \bar{1} \\ 4 & 1 & \bar{1} \\ 2 & 0 & 1 \end{bmatrix}$ | $\begin{bmatrix} 0.135\ldots & -0.866\ldots & 0.481\ldots \\ 0.234\ldots & 0.5 & 0.833\ldots \\ -0.962\ldots & 0 & 0.271\ldots \end{bmatrix}$ | (1 $\bar{1}$ $\bar{2}$), (0 0 $\bar{1}$) |

Here we show another example where this package can be applied to make an three-dimensional CSL interface by directly using the result from a diffraction experiment. Londoño-Calderon, A. *et al.*[26] has identified a Tellurium twinning GB by the diffraction pattern showing that two grains formed this GB comprising the $(11\bar{2}\bar{2})$ and $(\bar{2}11\bar{2})$ planes of the conventional hexagonal lattice, and that they have their $[1\bar{1}00]$ and $[0\bar{1}10]$ directions to be collinear. Our package can help to compute the corresponding target disorientation ($R^t$) by input the three-element miller indices of these two crystal planes and the indices of the two collinear directions, and a cif file of a conventional Tellurium unit cell. After determining this disorientation, this twinning GB can be constructed following the same routine for the last example of the CuInSe$_2$ twinning GB. Table 2 shows the involved parameters. Also, as shown in Figure 2, one can change the chirality of one of the crystals by the command:

my_interface.atoms_2 = [1,1,1] – my_interface.atoms_2

This should be down before building this interface. In this way, one can investigate the effects of the chirality of the two crystals on the physical properties of this twinning grain boundary.

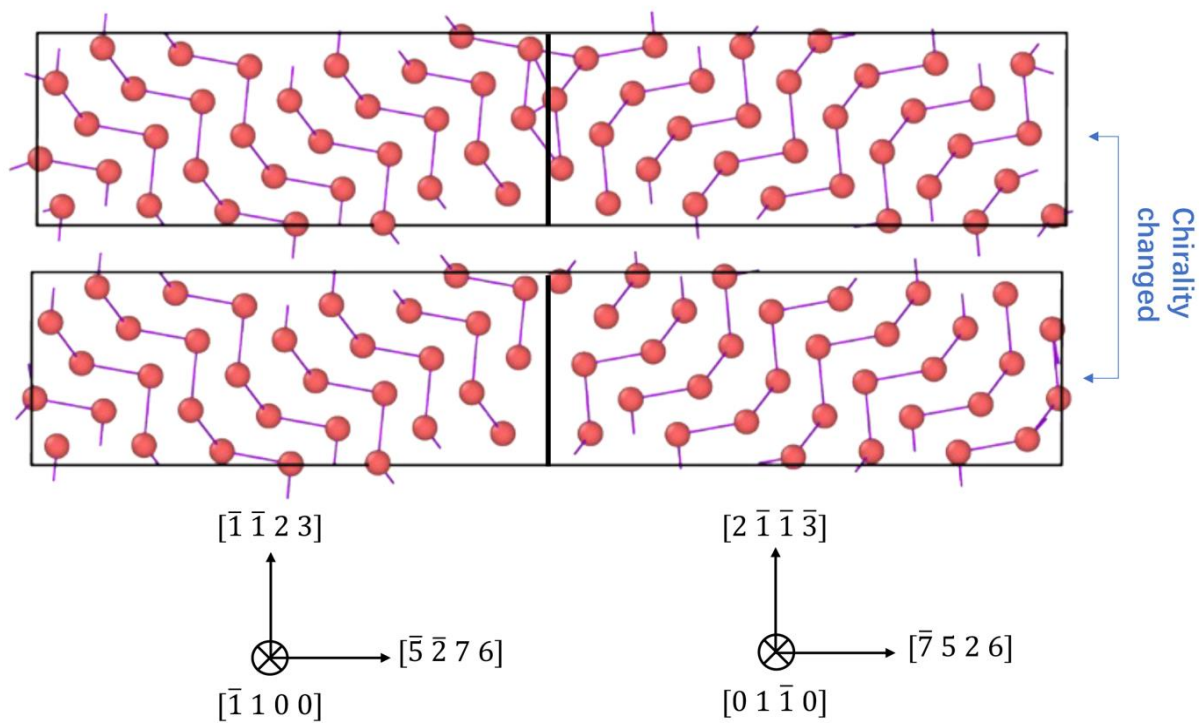

Figure 2 Bicrystals of the Tellurium twinning GBs with difference in chirality

### 3.2 Generating two-dimensional CSL interfaces – $\beta$-Si$_3$N$_4$(0001)/Si(111)

In **2.1** we explained the process to find a two-dimensional CSL lattice when a three-dimensional CSL lattice does not exist but a two-dimensional CSL lattice appears. Here we show an example of this case, which is the interface $\beta$-Si$_3$N$_4$(0001)/Si(111) reported in [27], as shown in Figure 3. The involved parameters making this interface is shown in Table 3.

Using this code, users can also adjust the termination of the two slabs at the interface as well as adjust the size of each slab. They can also adjust the vacuum size between the two interfaces and that at the end of the supercell. We also provide functions to make supercells sampling the RBT in the CNID. With the supercells generated by these operations, one can flexibly conduct effective high-throughput simulation for any conceptual or experimentally observed interface.

Table 3 Parameters involved generating the $\beta$-Si$_3$N$_4$(0001)/Si(111) interface

| $a_i^1$ (Si) | $a_i^{2,0}$ ($\beta$-Si$_3$N$_4$) | $R^t$ | $U_1$ $U_2$ |
|---|---|---|---|
| 2.74736 Å $\begin{bmatrix} \bar{1} & \bar{1} & 0 \\ \bar{1} & 0 & \bar{1} \\ 0 & \bar{1} & \bar{1} \end{bmatrix}$ | $\begin{bmatrix} 0 & \bar{a} & \bar{a} \\ 0 & \bar{b} & b \\ \bar{c} & 0 & 0 \end{bmatrix}$ $a = 3.83015$ Å $b = 6.63402$ Å $c = 2.92508$ Å | $\begin{bmatrix} -0.707\ldots & -0.408\ldots & 0.577\ldots \\ 0.707\ldots & -0.408\ldots & 0.577\ldots \\ -3.535\ldots & 0.816\ldots & 0.577\ldots \end{bmatrix}$ | $\begin{bmatrix} 0 & \bar{2} & 0 \\ \bar{2} & 2 & 0 \\ 2 & 0 & 1 \end{bmatrix}$ $\begin{bmatrix} 0 & 0 & 0 \\ 1 & 0 & 0 \\ 1 & -1 & 1 \end{bmatrix}$ |
| $U_b^1$ | $U_b^2$ | $D$ | (h$_1$ j$_1$ k$_1$) (h$_2$ j$_2$ k$_2$) |
| $\begin{bmatrix} \bar{1} & 0 & \bar{2} \\ \bar{1} & \bar{2} & 2 \\ \bar{1} & 2 & 0 \end{bmatrix}$ | $\begin{bmatrix} \bar{1} & 0 & 0 \\ 0 & 1 & 0 \\ 0 & 1 & \bar{1} \end{bmatrix}$ | $\begin{bmatrix} 1.006\ldots & 0.003\ldots & -0.009\ldots \\ -0.009\ldots & 1.006\ldots & 0.003\ldots \\ 0.003\ldots & -0.009\ldots & 1.006\ldots \end{bmatrix}$ | (1 1 1) (0 0 1) |

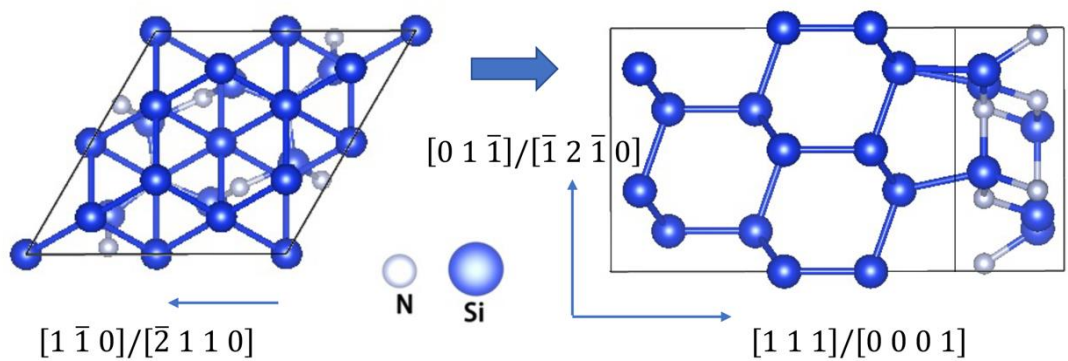

Figure 3 Bicrystal of the $\beta$-Si$_3$N$_4$(0001)/Si(111) interface

**Conclusion**

In this work, applying two effective algorithms we generated a python programme capable to build CSL and approximate CSL interfaces and compute their CNIDs. By inputting simply two cif files (better of conventional cells), and some intuitive parameters, users can build any desired near CSL interface. Our programme provides a convenient tool supporting in research on interfaces.

**Acknowledgements** Yaoshu Xie is supported by Ministry of Education, Culture,Sports, Science and Technology (MEXT) Scholarship with Embassy Recommendation for Chinese students. This study is supported by JST-CREST (JPMJCR1993) and the (MEXT); Nos 19H00818, and 19H05787, and Tenkai special funding in Institute of Industrial Science.